\documentclass[12pt]{article}
\usepackage{graphicx,amsmath,amssymb,amsfonts}
\usepackage{citesort,epsfig}

\newcommand{\DATE}  {\today}

\newcommand{\PPrtNo}
{
MSU-HEP-070101
}
\newcommand{\TITLE}
{
The Charm Parton Content of the Nucleon
}

\newcommand{\AUTHORS}
{ J.~Pumplin\footnote{E-mail: pumplin@msu.edu}$^a$,
H.L.~Lai$^{a,b,c}$, W.K.~Tung$^{a,b}$}
\newcommand{\INST}
{
$^a$ Michigan State University, E. Lansing, MI, USA \\
$^b$ University of Washington, Seattle, WA, USA \\
$^c$ Taipei Municipal University of Education, Taipei, Taiwan
}

\newcommand{\ABSTRACT}
{We investigate the charm sector of the nucleon structure phenomenologically,
using the most up-to-date global QCD analysis. Going beyond the common
assumption of purely radiatively generated charm, we explore possible degrees
of freedom in the parton parameter space associated with nonperturbative
(intrinsic) charm in the nucleon. Specifically, we explore the limits that
can be placed on the intrinsic charm (IC) component, using all relevant
hard-scattering data, according to scenarios in which the IC has a form
predicted by light-cone wave function models; or a form similar to the 
light sea-quark distributions. We find that the range of IC is constrained 
to be from zero (no IC) to a level 2--3 times larger than previous model 
estimates.  The behaviors of typical charm distributions within this range are 
described, and their implications for hadron collider phenomenology are 
briefly discussed.}

\newcommand{\figA}
{
\begin{figure}[htb]
 \centerline{
 \resizebox*{!}{0.3\textheight} {
 \includegraphics{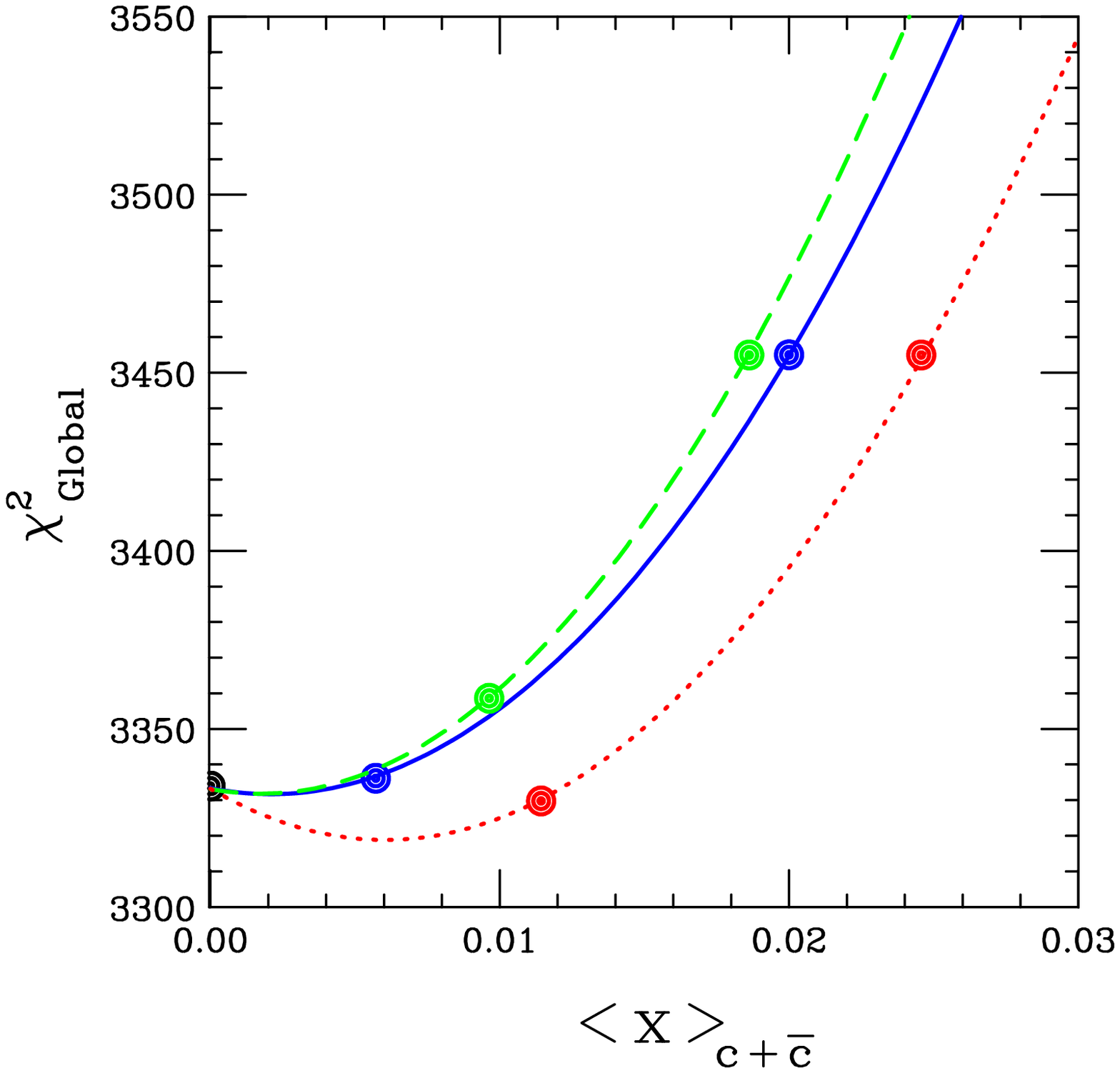}}
 }
\caption{ Goodness-of-fit vs.\ momentum fraction of IC at the starting 
scale $\mu=1.3 \, \mathrm{GeV}$ for three models of IC: 
BHPS (solid curve); 
meson cloud (dashed curve); and 
sea-like (dotted curve). Round dots indicate the specific
fits that are shown in Figs.~\ref{fig:figB}--\ref{fig:figD}.}
\label{fig:figA}
\end{figure}
}
\newcommand{\figB}
{
\begin{figure}[htb]
\resizebox{0.32\textwidth}{!}{
\includegraphics[clip=true,scale=0.32]{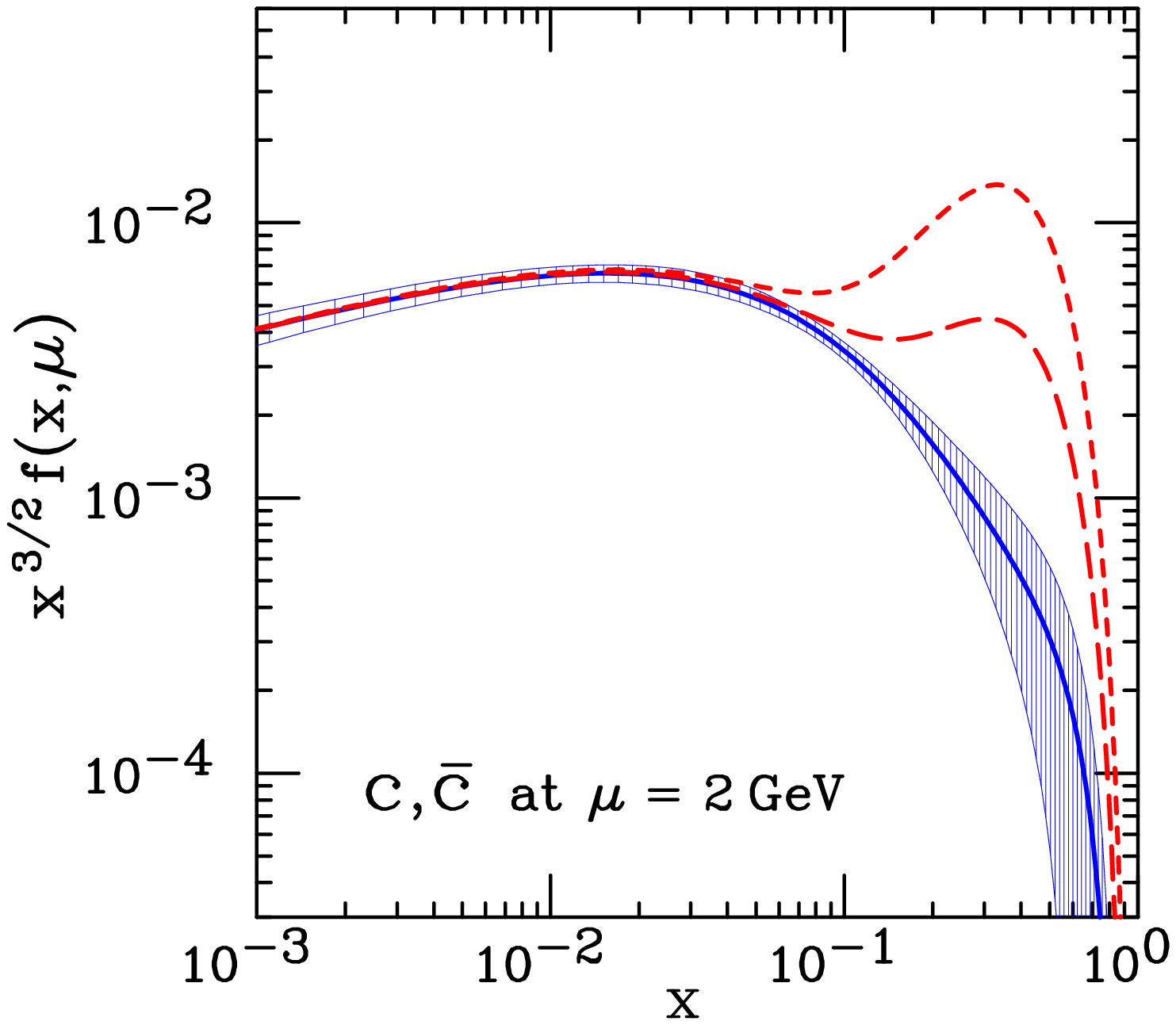}
}
\hfill
\resizebox{0.32\textwidth}{!}{
\includegraphics[clip=true,scale=0.32]{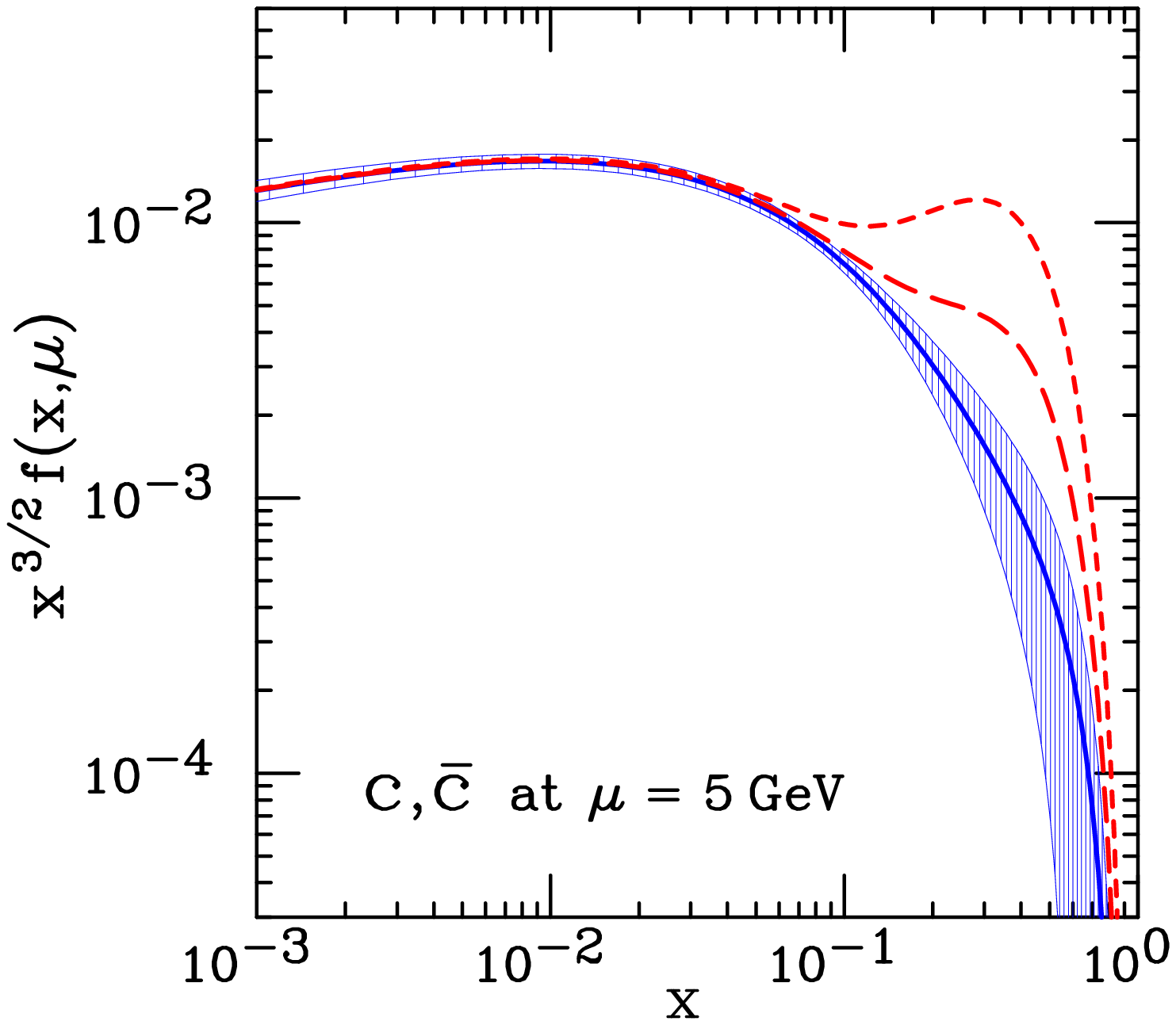}
}
\hfill
\resizebox{0.32\textwidth}{!}{
\includegraphics[clip=true,scale=0.32]{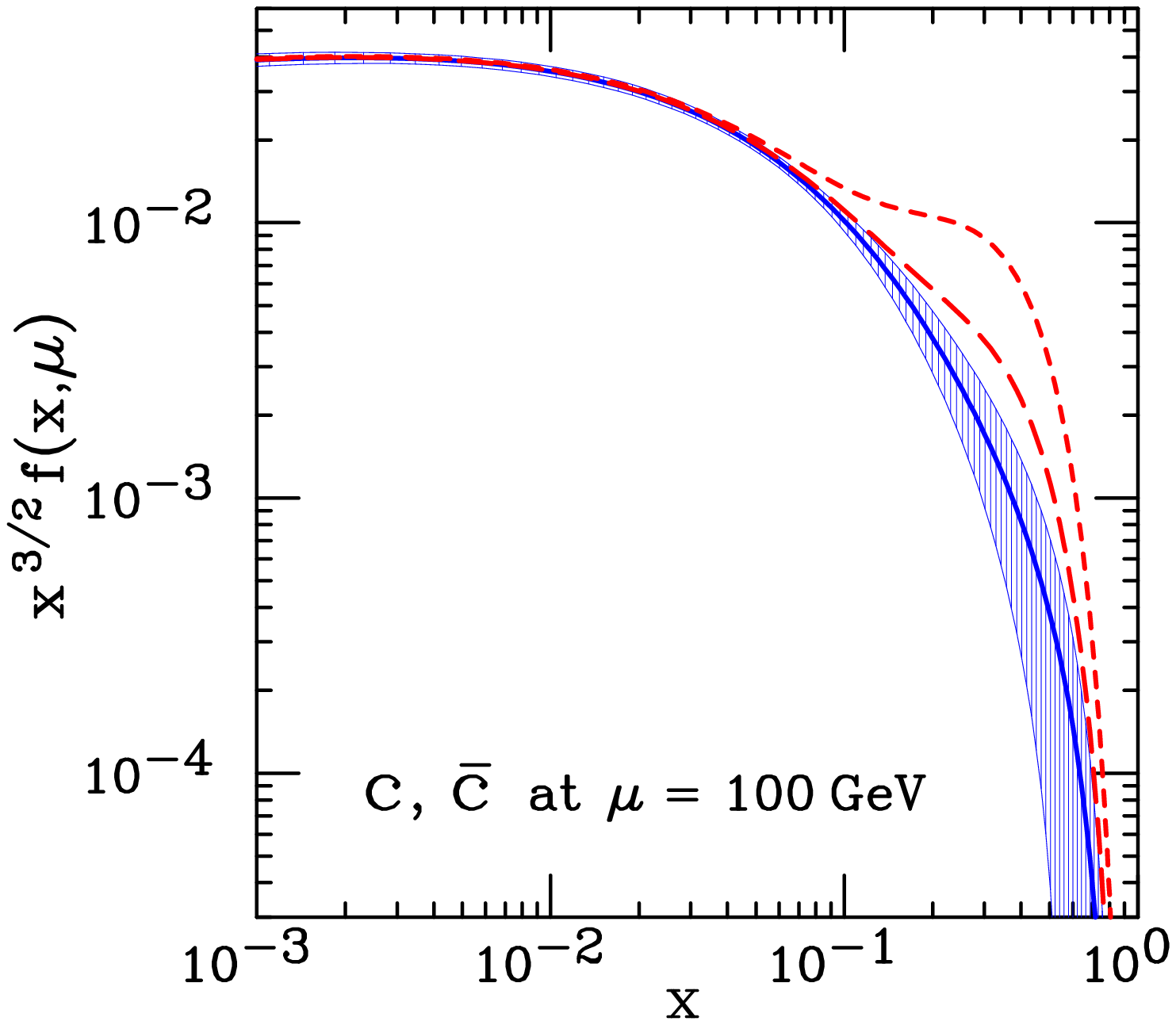}
} \caption{ Charm quark distributions from the BHPS IC model. The three
panels correspond to scales $\mu = 2$, $\mu = 5$, and $\mu = 100 \,
\mathrm{GeV}$. The long-dash (short-dash) curve corresponds to $\langle x
\rangle_{c + \bar{c}} = 0.57\%$ ($2.0\%$). The solid curve and shaded region
show the central value and uncertainty from CTEQ6.5, which contains
no IC.} \label{fig:figB}
\end{figure}
}
\newcommand{\figC}
{
\begin{figure}[htb]
\resizebox{0.32\textwidth}{!}{
\includegraphics[clip=true,scale=0.32]{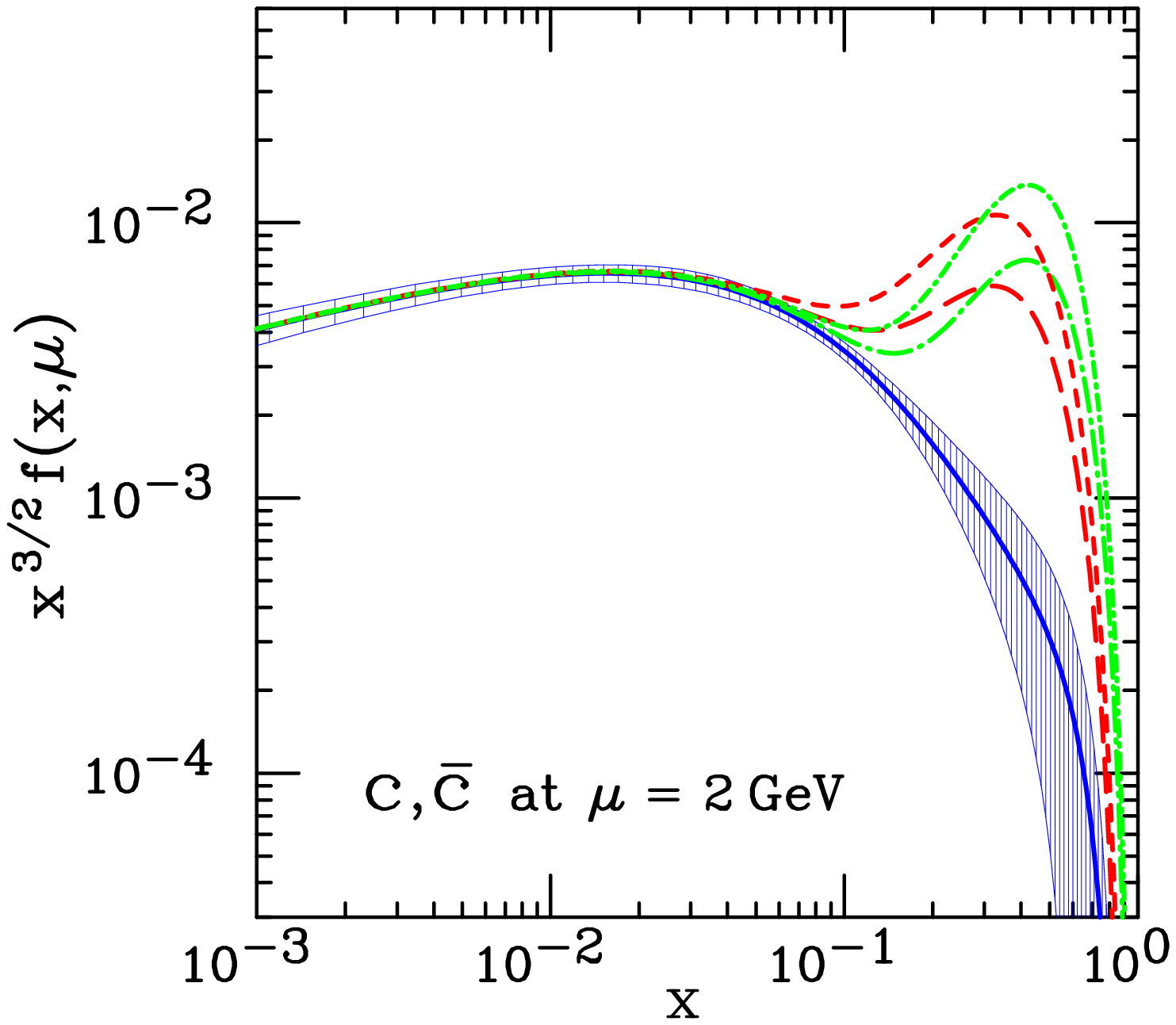}
}
\hfill
\resizebox{0.32\textwidth}{!}{
\includegraphics[clip=true,scale=0.32]{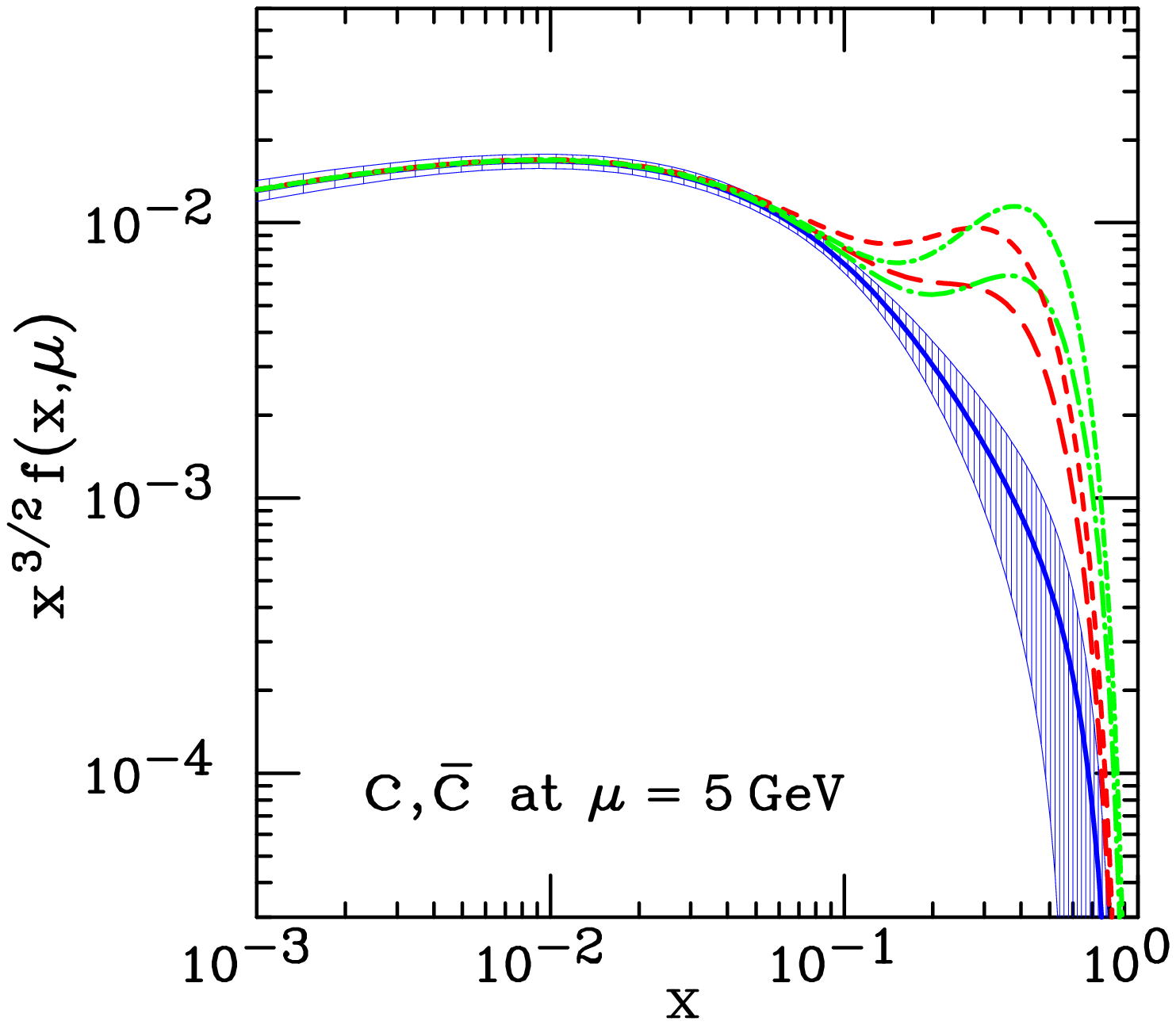}
}
\hfill
\resizebox{0.32\textwidth}{!}{
\includegraphics[clip=true,scale=0.32]{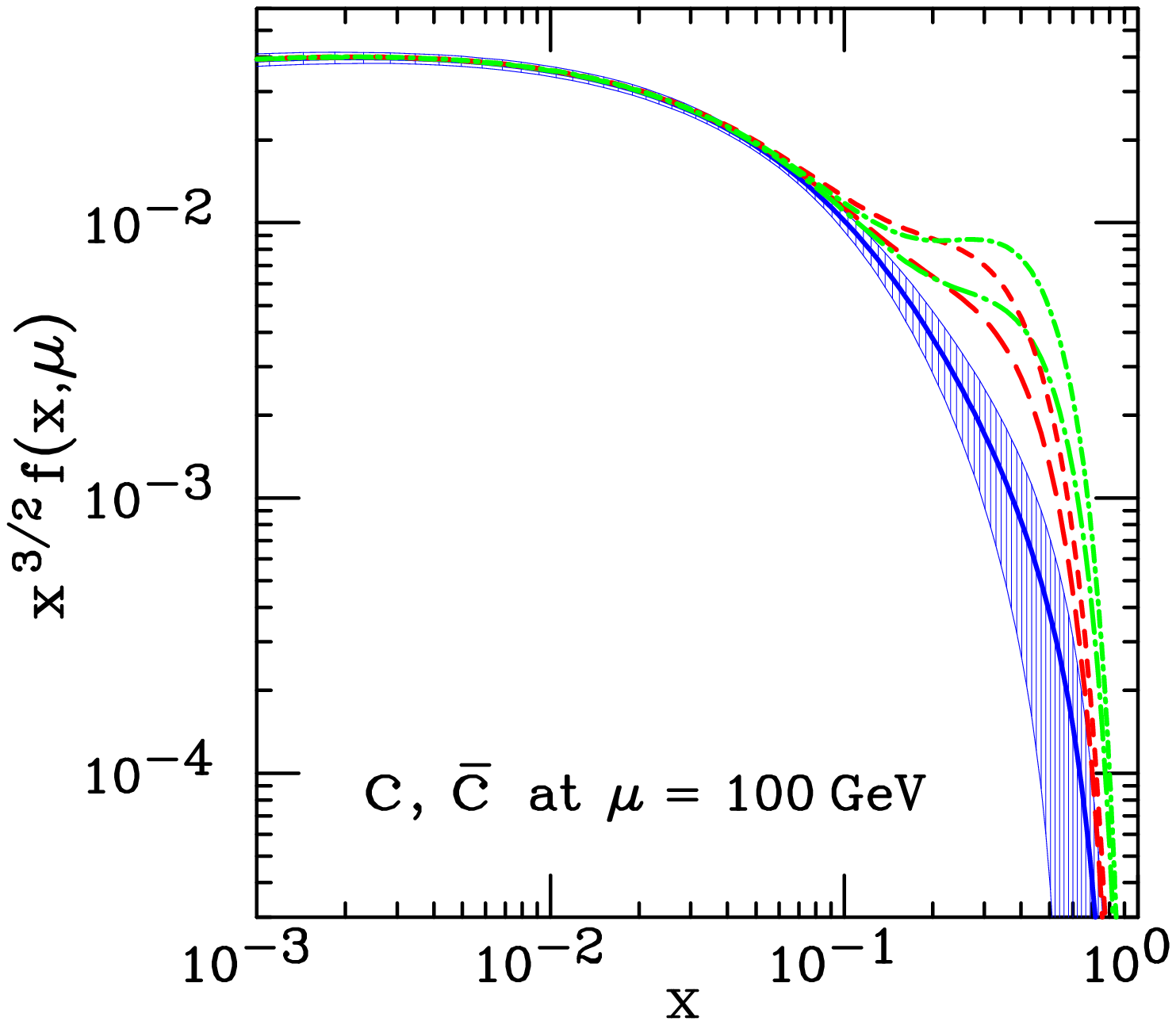}
} \caption{Same as Fig.\,\ref{fig:figB}, except for the meson
cloud model. The long-dash (short-dash) curves correspond to
$\langle x \rangle_{c + \bar{c}} = 0.96\%$ ($1.9\%$).}
\label{fig:figC}
\end{figure}
}
\newcommand{\figD}
{
\begin{figure}[htb]
\resizebox{0.32\textwidth}{!}{
\includegraphics[clip=true,scale=0.32]{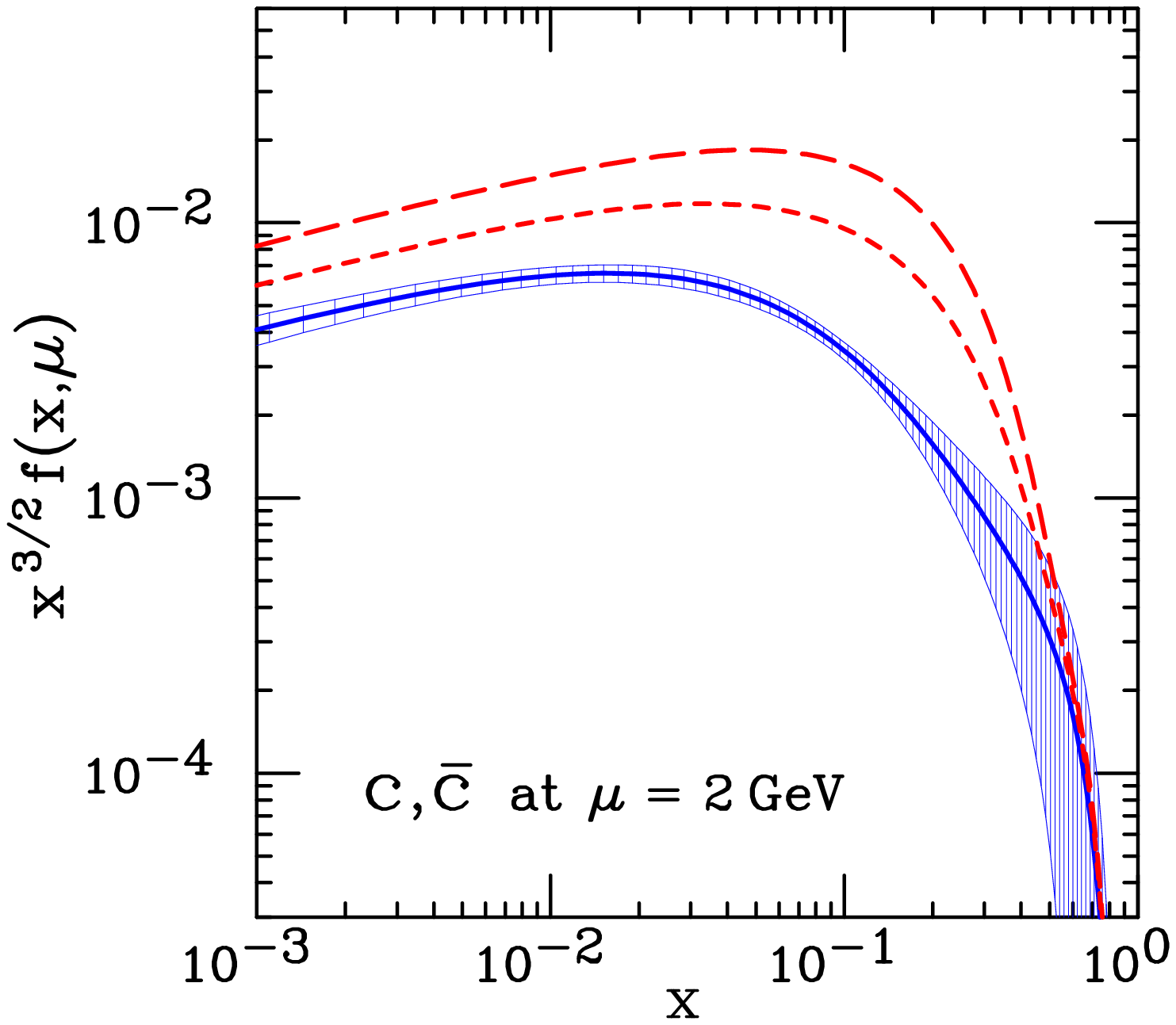}
}
\hfill
\resizebox{0.32\textwidth}{!}{
\includegraphics[clip=true,scale=0.32]{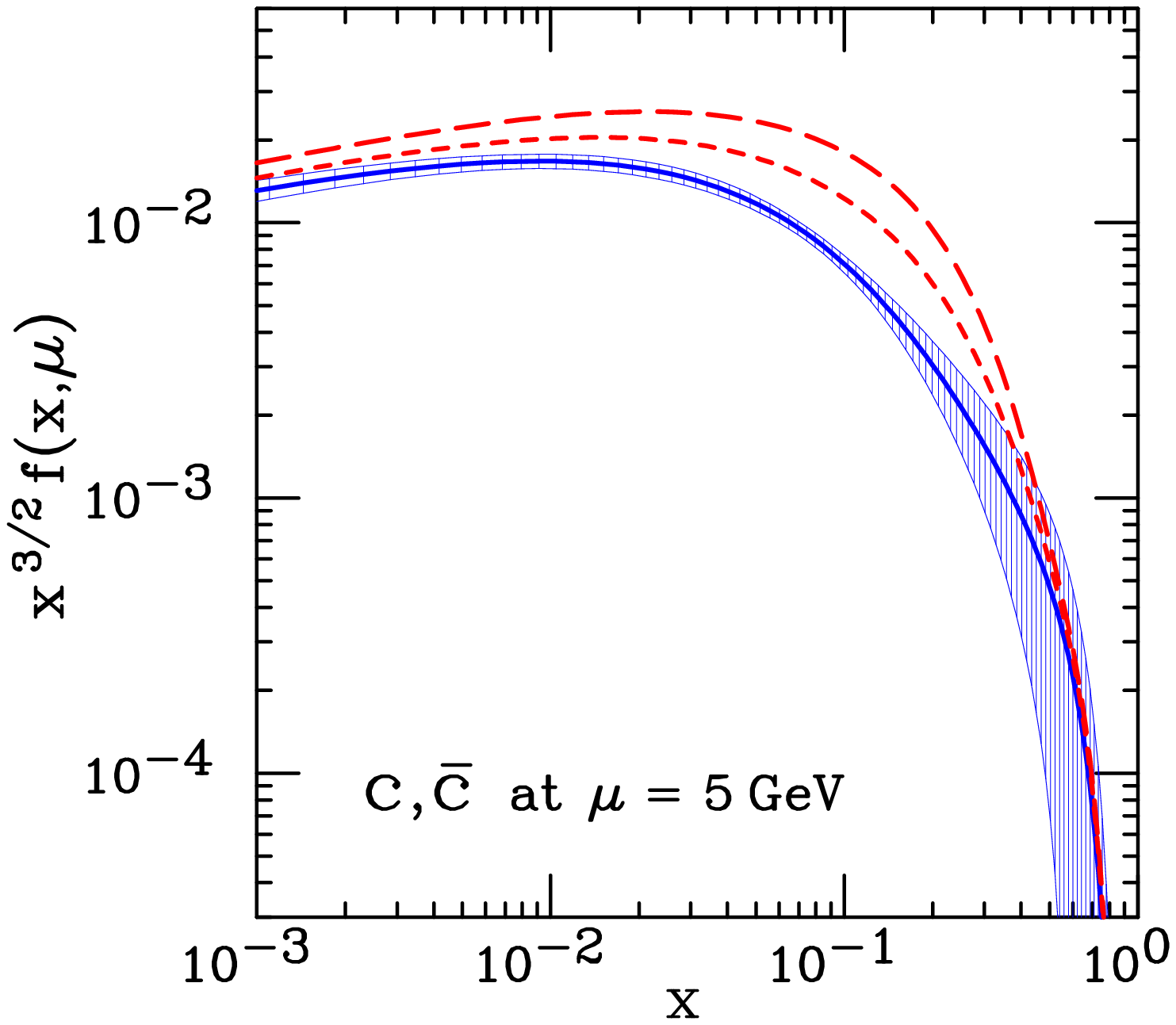}
}
\hfill
\resizebox{0.32\textwidth}{!}{
\includegraphics[clip=true,scale=0.32]{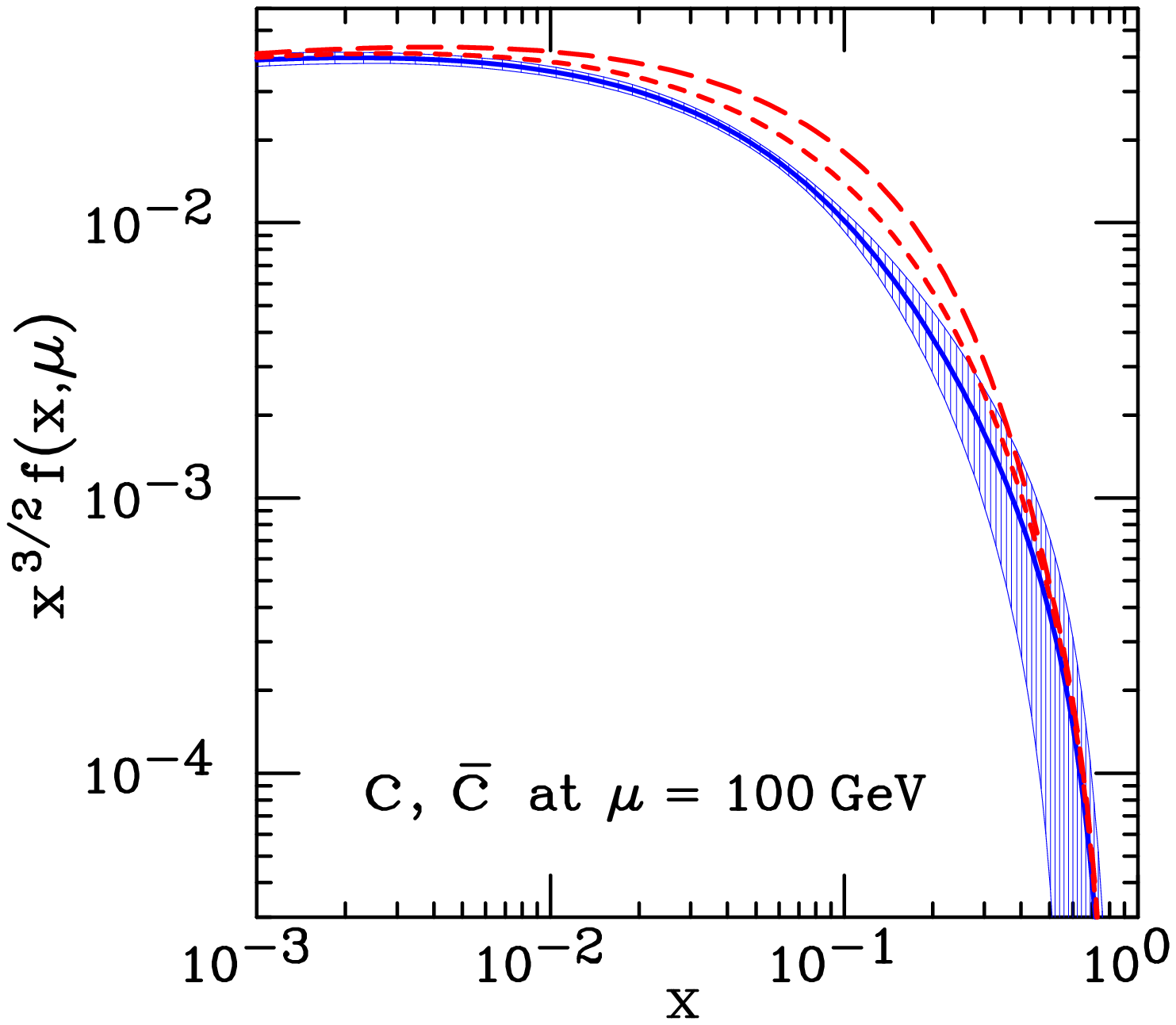}
} \caption{ Same as Fig.\,\ref{fig:figB}, except for the sea-like
scenario. The long-dash (short-dash) curves correspond to $\langle x
\rangle_{c + \bar{c}} = 2.4\%$ ($1.1\%$). } \label{fig:figD}
\end{figure}
}
\newcommand{\figCOMPcx}
{
\begin{figure}[htb]
\resizebox{0.32\textwidth}{!}{
\includegraphics[clip=true,scale=0.32]{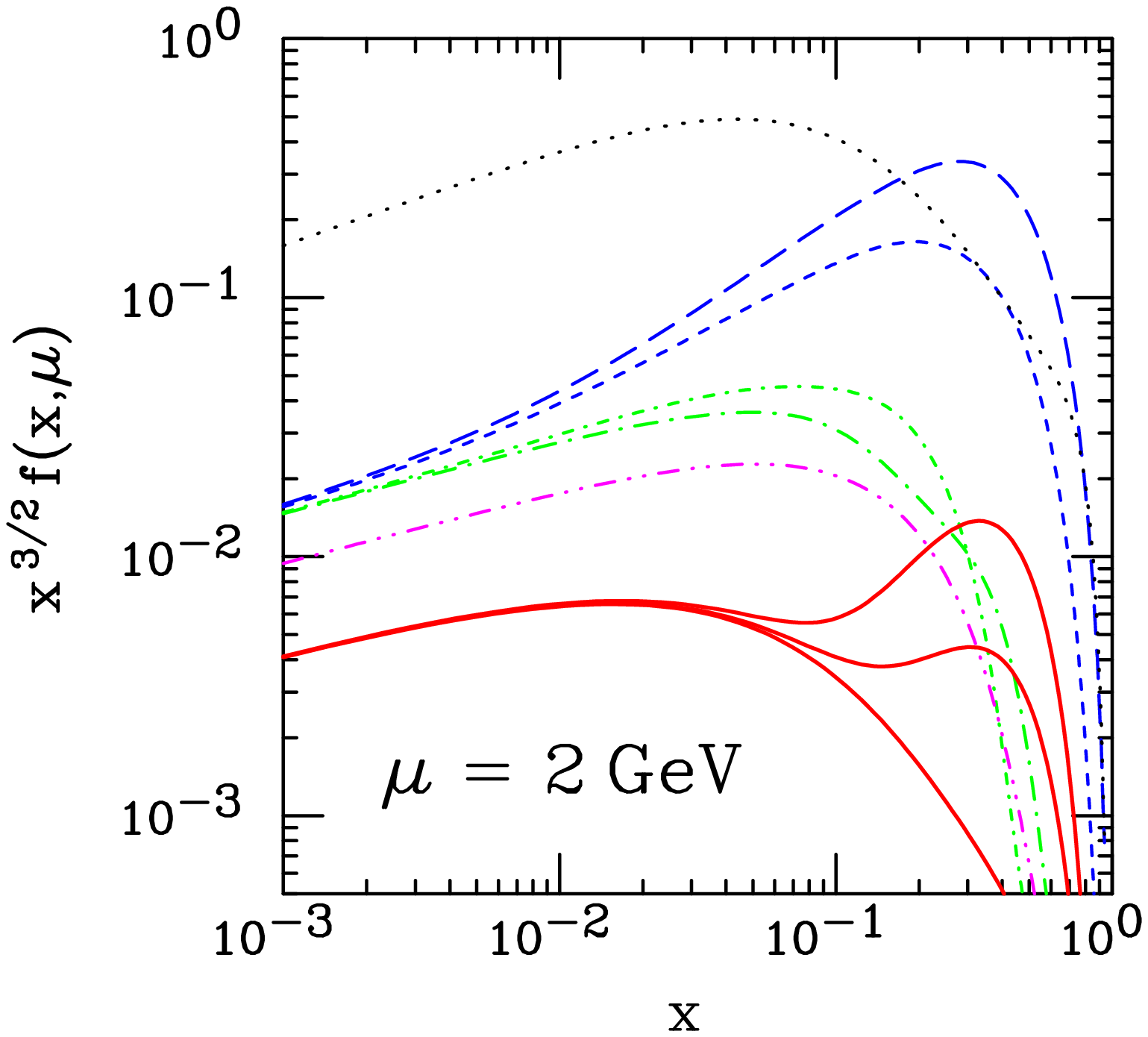}
}
\hfill
\resizebox{0.32\textwidth}{!}{
\includegraphics[clip=true,scale=0.32]{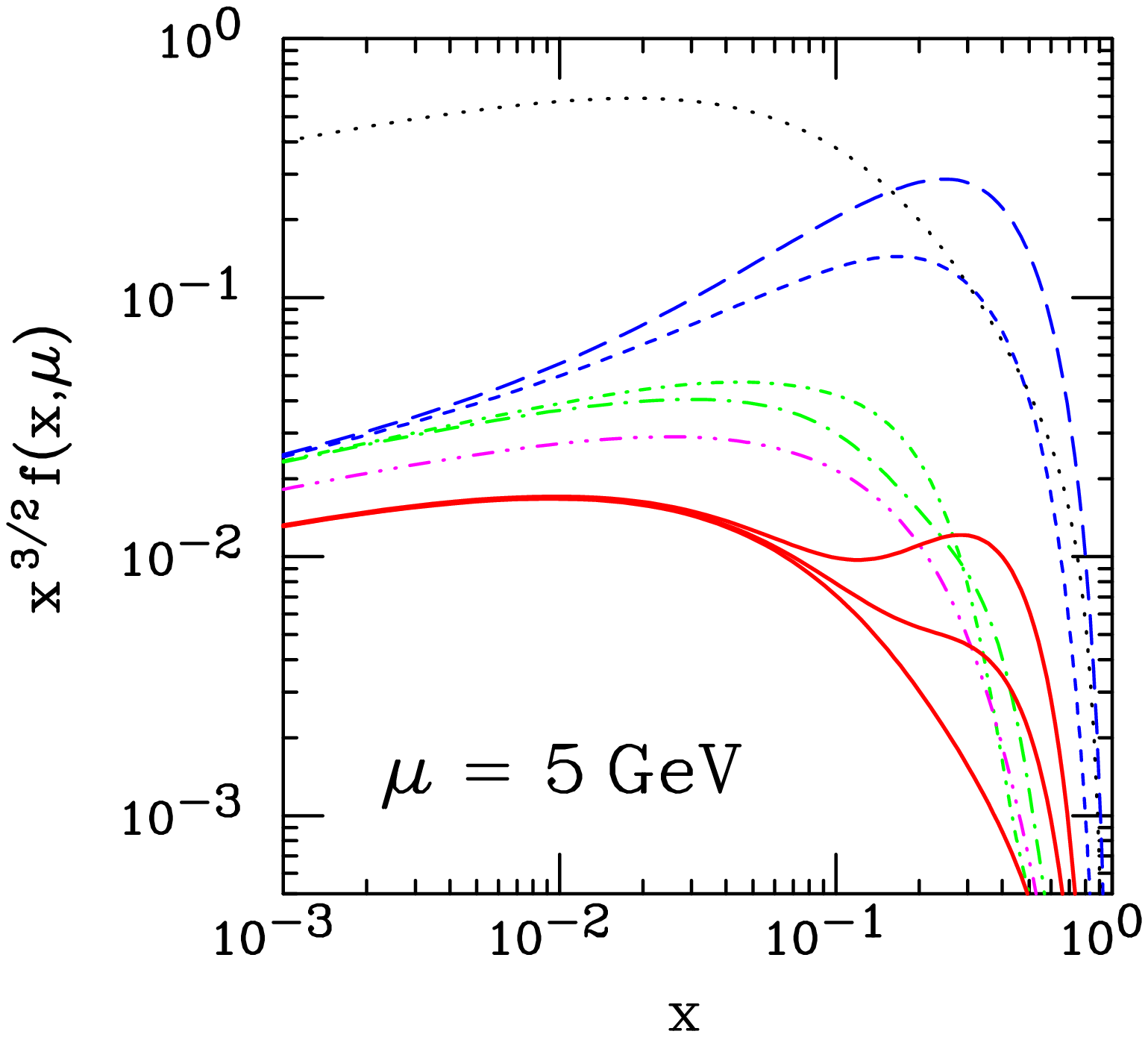}
}
\hfill
\resizebox{0.32\textwidth}{!}{
\includegraphics[clip=true,scale=0.32]{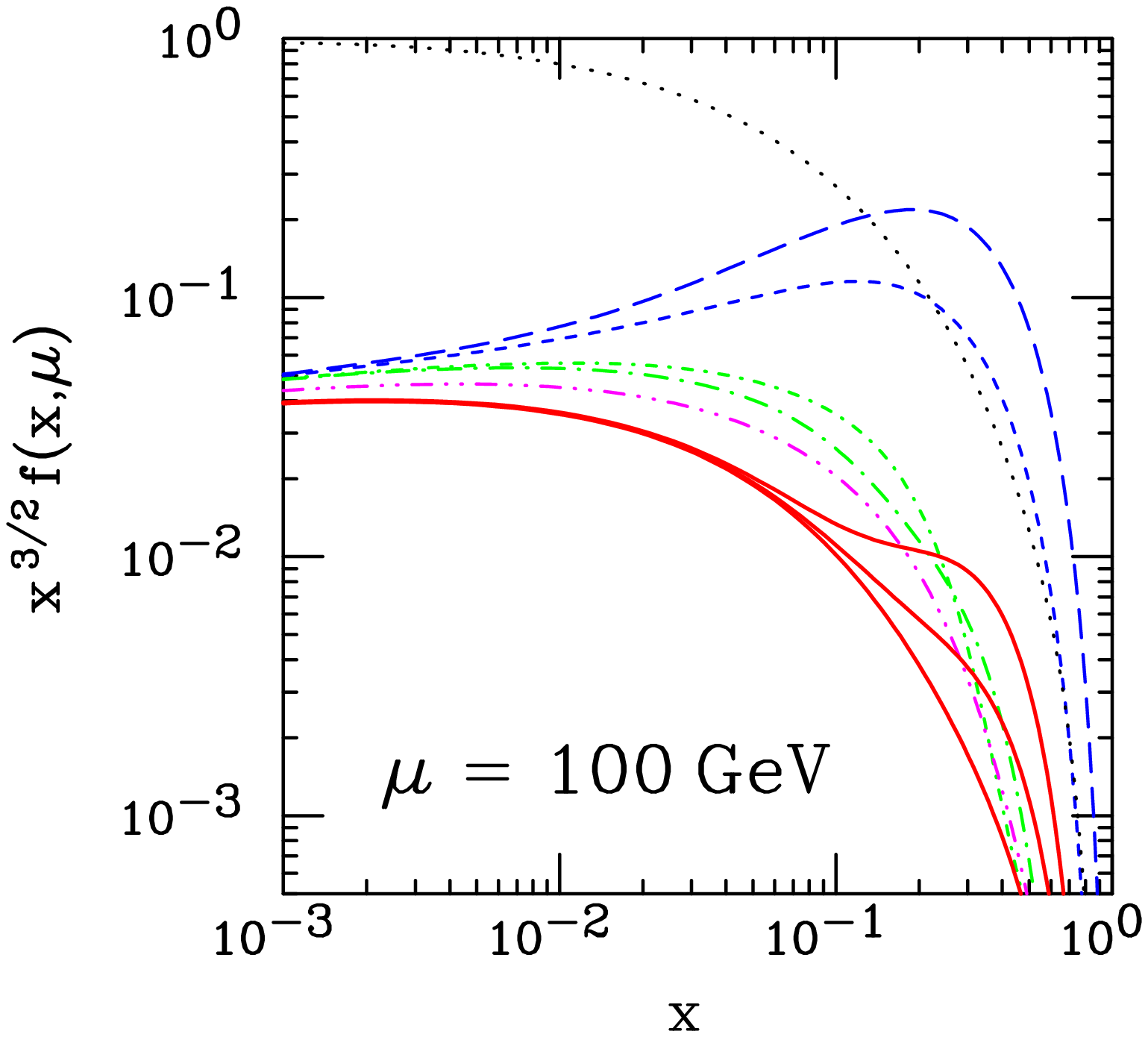}
} \caption{Comparison of charm with other flavors:
$u, \bar{u}$ (long-dash, long-dash-dot);
$d, \bar{d}$ (short-dash, short-dash-dot);
$s = \bar{s}$ (dash-dot-dot), $g$ (dot).
The solid curves are $c = \bar{c}$ with no IC (lowest) or the two
magnitudes of IC in the BHPS model that are discussed in the text.
}
\label{fig:figCOMPcx}
\end{figure}
}
\newcommand{\figCOMPca}
{
\begin{figure}[htb]
\resizebox{0.32\textwidth}{!}{
\includegraphics[clip=true,scale=0.32]{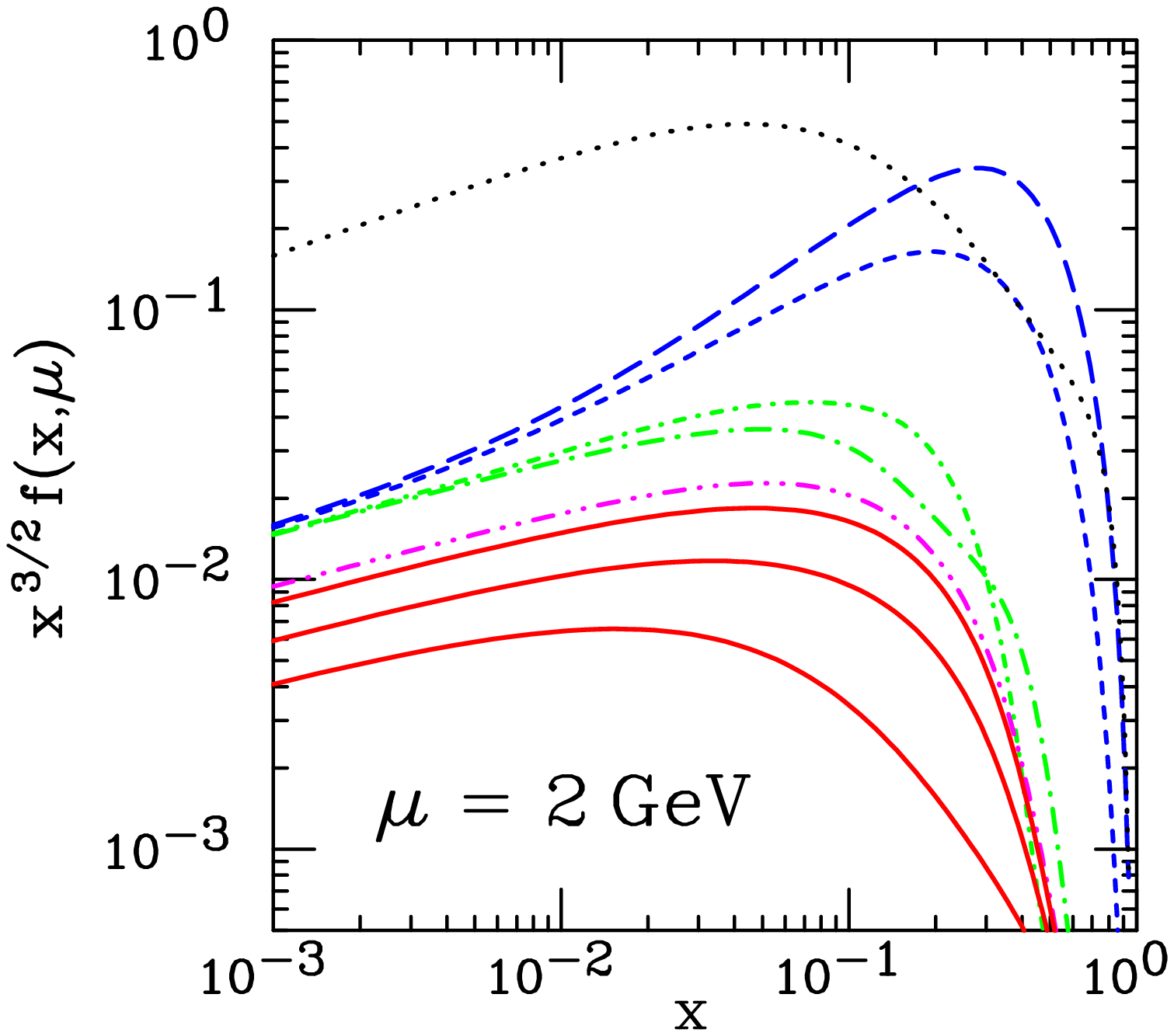}
}
\hfill
\resizebox{0.32\textwidth}{!}{
\includegraphics[clip=true,scale=0.32]{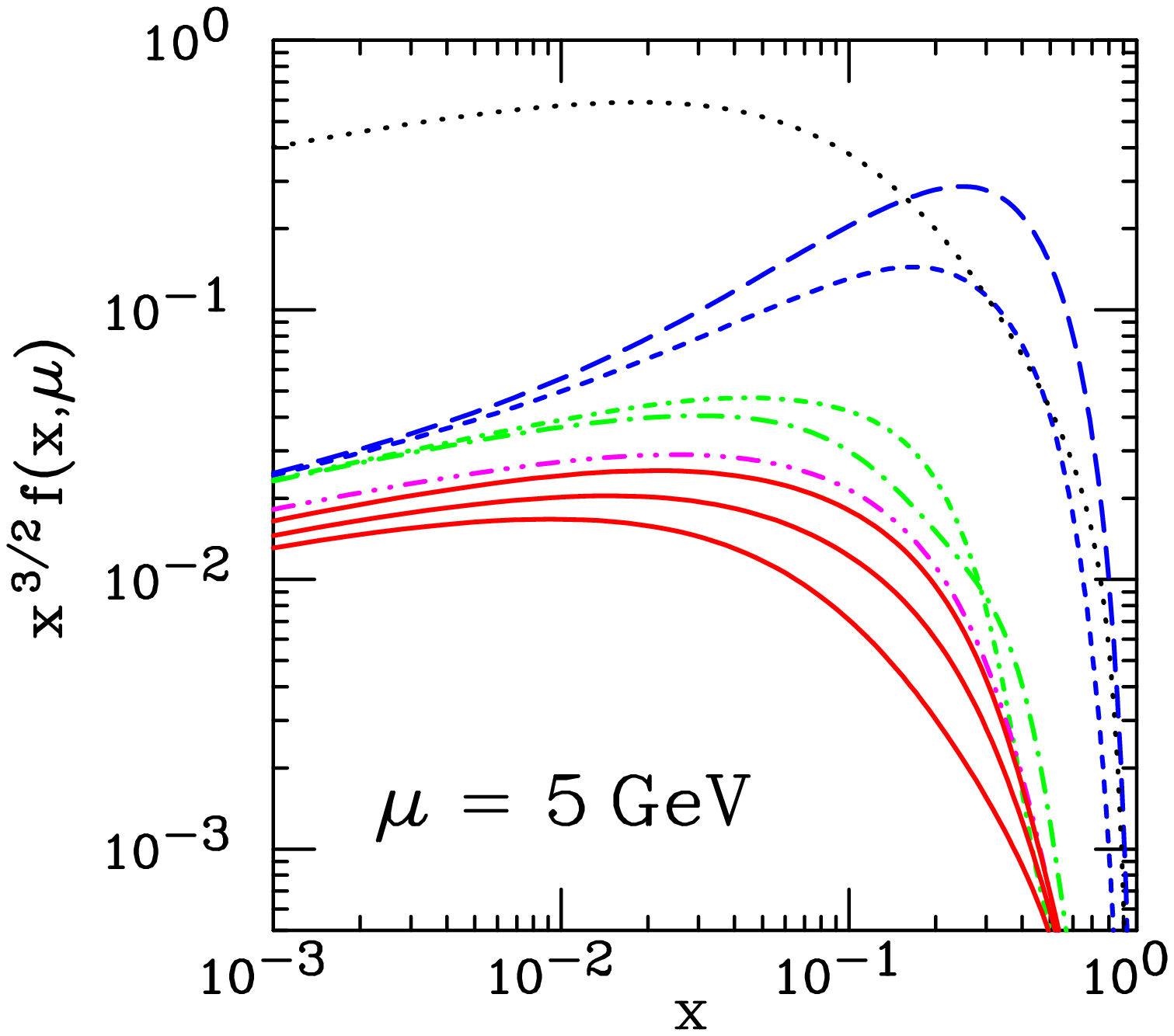}
}
\hfill
\resizebox{0.32\textwidth}{!}{
\includegraphics[clip=true,scale=0.32]{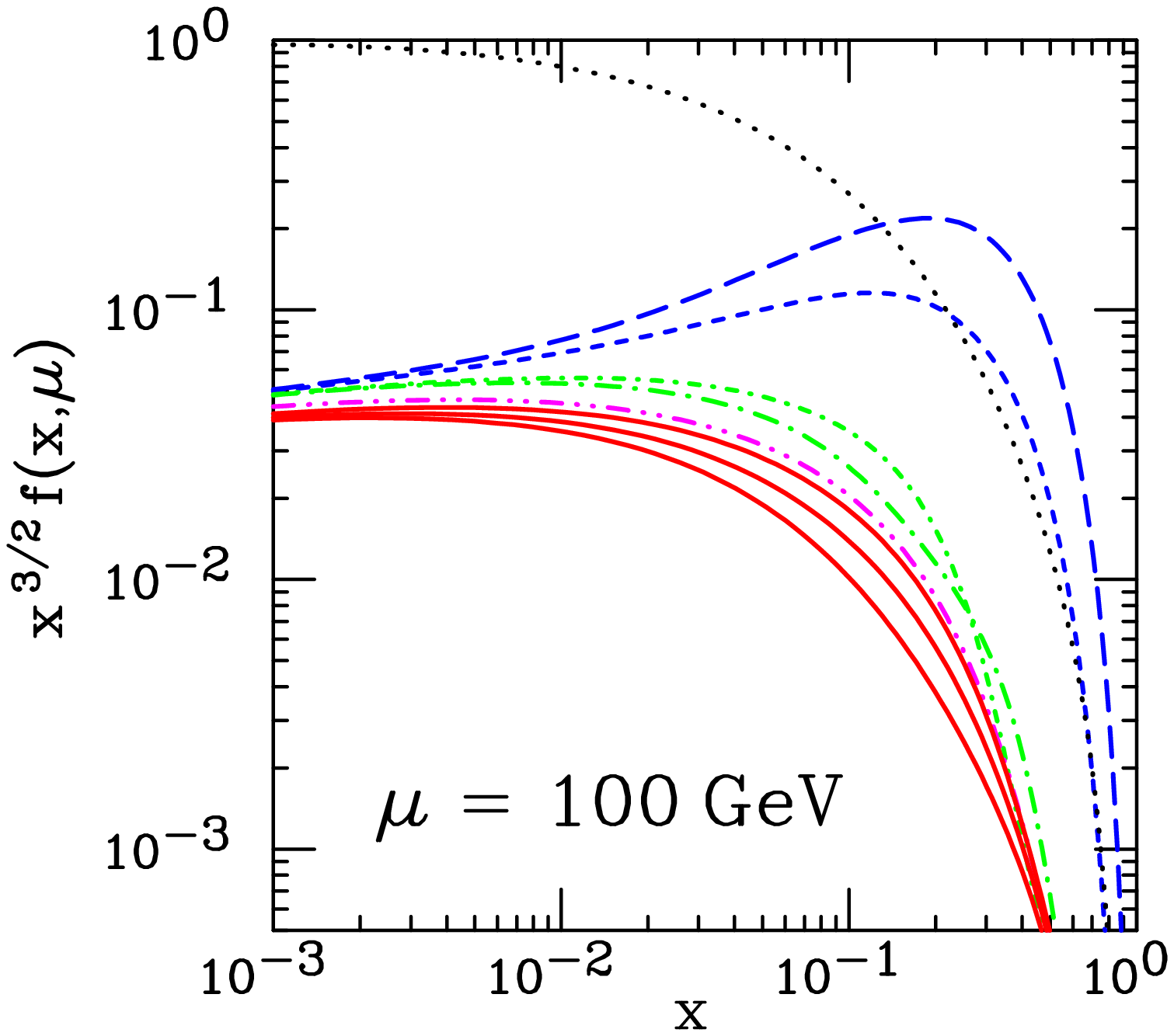}
} \caption{ Same as Fig.\,\ref{fig:figCOMPcx}, but for sea-like IC. }
\label{fig:figCOMPca}
\end{figure}
}
\newcommand{\figE}
{
\begin{figure}[htb]
\resizebox{0.32\textwidth}{!}{
\includegraphics[clip=true,scale=0.32]{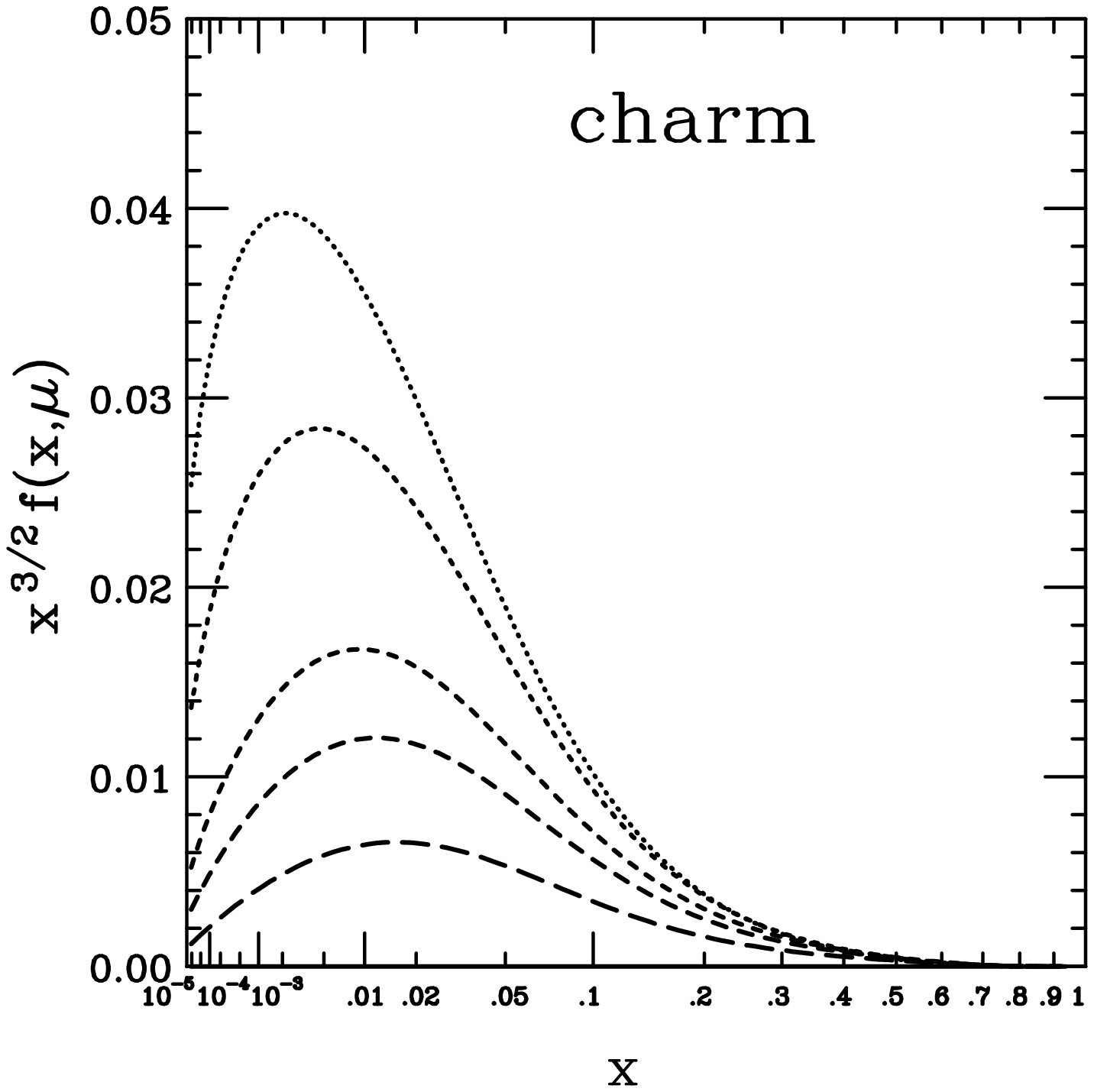}
}
\hfill
\resizebox{0.32\textwidth}{!}{
\includegraphics[clip=true,scale=0.32]{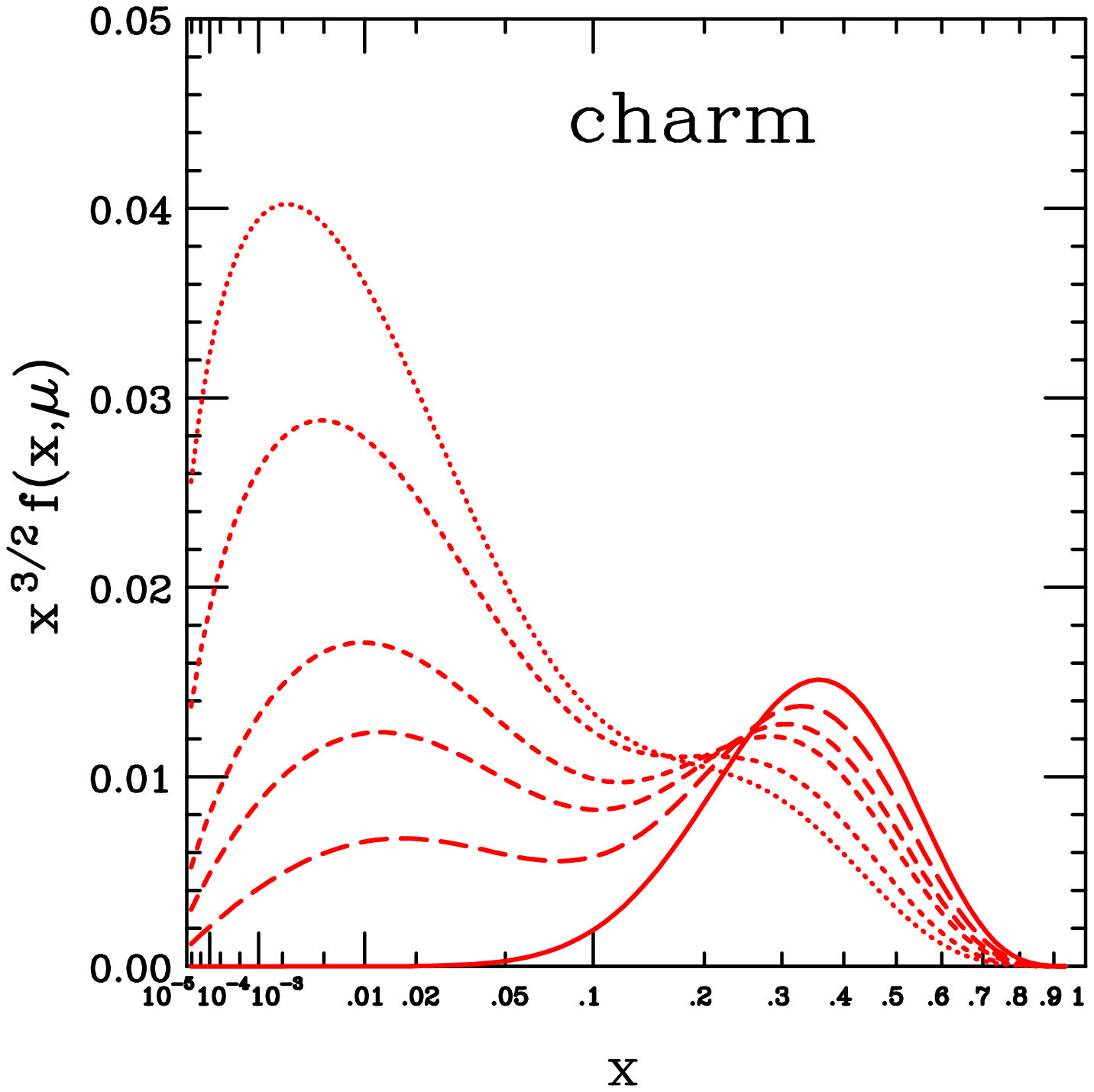}
}
\hfill
\resizebox{0.32\textwidth}{!}{
\includegraphics[clip=true,scale=0.32]{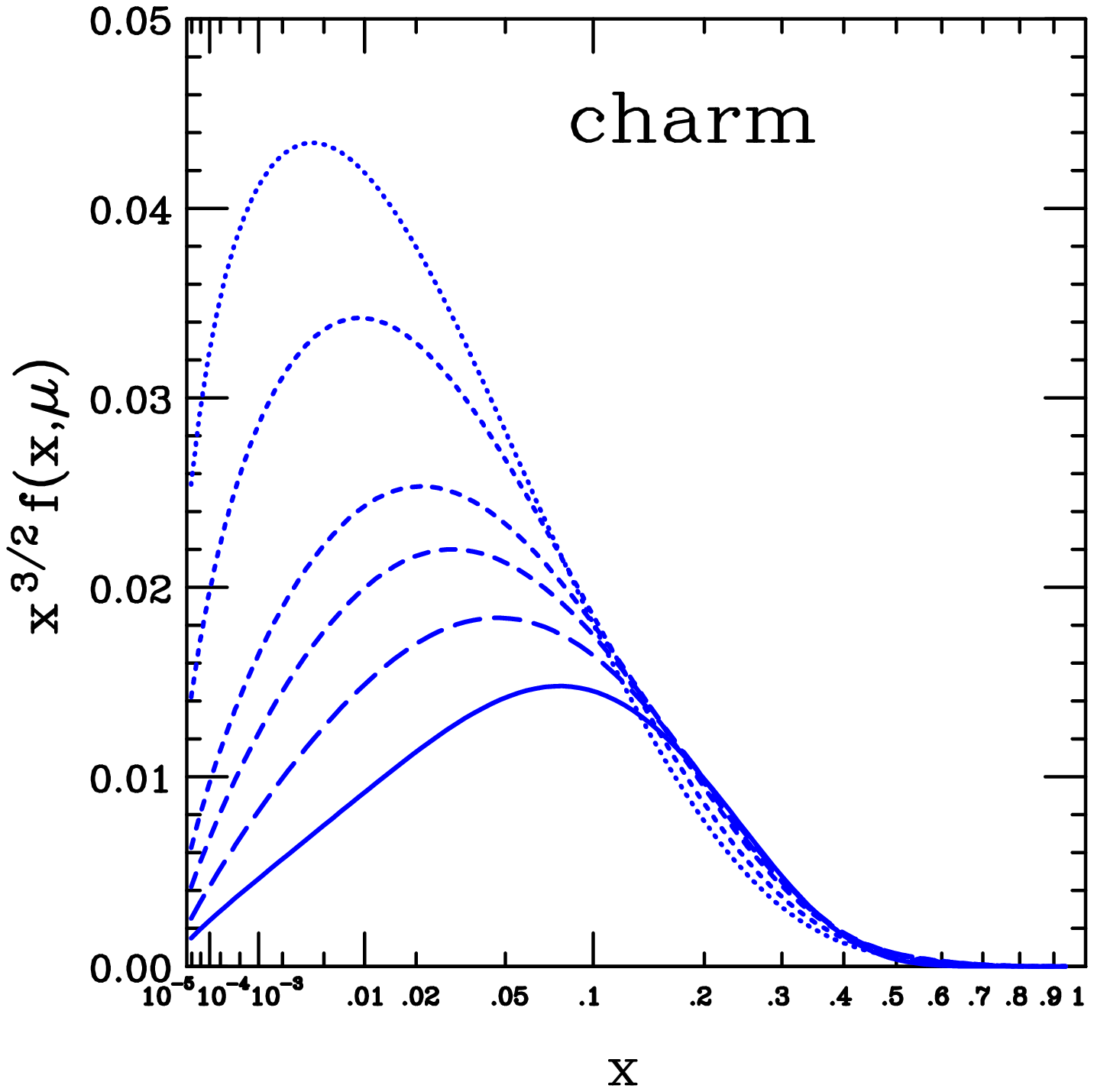}
}
\caption{
Charm distributions at scale $\mu = 1.3$ (solid), $2.0$, $3.16$, $5$, $20$,
$100 \, \mathrm{GeV}$ (dotted).
Left: no IC;
center: BHPS model with maximal IC consistent with experiment;
right: sea-like model with maximal IC consistent with experiment.
}
\label{fig:figE}
\end{figure}
}

\begin{document}

% Standard Simple Cover Page for Papers -- wkt 1/11/95

\begin{titlepage}

\noindent
\begin{tabular}{l} \DATE
\end{tabular}
\hfill
\begin{tabular}{l}
\PPrtNo
\end{tabular}

\vspace{1cm}

\begin{center}
                           % Title
\renewcommand{\thefootnote}{\fnsymbol{footnote}}
{
\LARGE \TITLE
%\footnote[2]{\THANKS}
}

\vspace{1.25cm}
                          % Authors
{\large  \AUTHORS}

\vspace{1.25cm}

                          % Institutions
\INST
\end{center}

\vfill

\ABSTRACT                 % Abstract

\vfill

\newpage
\end{titlepage}

\setcounter{footnote}{0}
\tableofcontents
\newpage

%TCIDATA{Version=4.10.0.2363}
%TCIDATA{LaTeXparent=0,0,../charm.tex}

%TCIDATA{ChildDefaults=chapter:1,page:1}

\section{Introduction}

\label{sec:Introduction}

The parton distribution functions (PDFs) that describe the quark and gluon
structure of the nucleon at short distances are essential inputs to the
calculation of all high energy processes at hadron colliders. They are
therefore important for carrying out searches for New Physics, as well as for
making precision tests of the Standard Model (SM), in regions where
perturbative Quantum Chromodynamics (PQCD) theory is applicable.  Although
much progress has been made in the last twenty years in determining the PDFs
by global QCD analysis of a wide range of hard processes, very little is
known phenomenologically about the heavy quark (charm $c$ and bottom $b$) and
antiquark ($\bar{c}$, $\bar{b}$) content of the
nucleon \cite{wkt04HCP,Thorne:2006wq}.

Knowledge of the heavy quark components is inherently important as an aspect 
of the fundamental structure of the nucleon. In addition, the heavy quarks 
are expected to play an increasingly significant role in the physics programs 
of the Tevatron Run II and the Large Hadron Collider (LHC), since many new
processes of interest, such as single-top production and Higgs
production in the SM and beyond, are quite sensitive to the heavy quark
content of the nucleon.

Existing global QCD analyses extract the PDFs by comparing a wide range of
hard-scattering data to perturbative QCD theory. In these analyses, one
usually adopts the ansatz that heavy quark partons in the nucleon are
\textquotedblleft radiatively generated\textquotedblright, i.e., they
originate only from QCD evolution, starting from a null distribution at a
factorization scale of approximately the relevant quark mass. This is
motivated, on the theoretical side, by the notion that heavy quark degrees of
freedom should be perturbatively calculable; and on the practical side, by
the lack of clearly identifiable experimental constraints on these degrees of
freedom in existing data. Neither of those considerations justifies the
ansatz, however---especially for charm, whose mass lies in between the soft
and hard energy scales. In fact, many nonperturbative models, particularly
those based on the light-cone wave function picture, expect an
\textquotedblleft intrinsic charm\textquotedblright\ (IC) component of the
nucleon at an energy scale comparable to $m_{c}$, the mass of the charm
quark. This IC component, if present at a low energy scale, will participate
fully in QCD dynamics and evolve along with the other partons as the energy
scale increases. It can therefore have observable consequences on physically
interesting processes at high energies and short distances.

With recent advances in the implementation of the general perturbative QCD
formalism to incorporate heavy quark mass effects \cite{MrstHq,cteq65m}, and
the availability of comprehensive precision data from HERA, the Tevatron,
and fixed-target experiments, we are now in a position to 
study the charm content of the nucleon phenomenologically,
with minimal model-dependent assumptions. This paper
represents a first systematic effort to perform this study and
answer the following questions: (i) do current theory and experiment
determine, or place useful limits on, the charm component of the nucleon at
the scale of $m_c$; (ii) if a non-vanishing charm distribution is allowed, 
can current global QCD analysis distinguish its shape between a form typical 
of light sea quarks (peaked at small $x$) and the form predicted in light-cone
wave function models (concentrated at moderate and large $x$); and (iii) what
are the implications of IC for the Tevatron and the LHC physics programs?

%TCIDATA{Version=4.10.0.2363}
%TCIDATA{LaTeXparent=0,0,../charm.tex}

%TCIDATA{ChildDefaults=chapter:2,page:1}

\section{Charm Partons at the Scale $\mathbf{\protect\mu_{0}\approx m_{c}}$}

\label{sec:IntrinsicCharm}

Let $f_{a}(x,\mu )$ denote the PDF of parton flavor $a$ inside the proton at
momentum fraction $x$ and factorization scale $\mu $. At short
distances, corresponding to large $\mu $, the scale-dependence of $%
f_{a}(x,\mu)$ is governed by the QCD evolution equation, with perturbatively
calculable evolution kernals (splitting functions). Thus, the set of PDFs $%
\left\{ f_{a}(x,\mu )\right\} $ are fully determined once their functional
form in $x$ is specified at a fixed scale $\mu =\mu _{0}$, provided $\mu _{0}
$ is large enough to be in the region where PQCD applies. In practice, $\mu
_{0}$ is usually chosen to be on the order $1-2\, \mathrm{GeV}$, which
is at the borderline between the short-distance (perturbative) and
long-distance (nonperturbative) regions. For the gluon and the light quarks (%
$a=g,u,d,s$), the $f_{a}(x,\mu _{0})$ are certainly nonperturbative in origin.
They must be determined phenomenologically through global QCD analysis that
compares the theoretical predictions with a wide range of experimental data
on hard processes \cite{cteq6,mrst,cteq65m}. In the case of charm and bottom
quarks, we need to examine the situation more closely.  In this paper, 
we shall focus in particular on charm. For convenience, we use the short-hand 
notation $c(x,\mu)\equiv f_c(x,\mu)$, and frequently omit the argument $\mu$, 
i.e.~use $c(x)$ in place of $c(x,\mu)$.

Over the energy range of most interest to current high energy physics, the
charm quark behaves as a parton, and is characterized by a PDF $c(x,\mu )$
that is defined for $\mu \geqslant m_{c}$. It is common in global QCD
analysis to consider charm a \emph{heavy quark}, and to adopt the ansatz $%
c(x,\mu _{0}=m_{c})=0$ as the initial condition for calculating $c(x,\mu )$
at higher energy scales by QCD evolution. This is the so-called \emph{%
radiatively generated charm} scenario. In the context of global analysis,
this ansatz implies that the charm parton does not have any independent
degrees of freedom in the parton parameter space: $c(x,\mu )$ is completely
determined by the gluon and light quark parton parameters.

However, nature does not have to subscribe to this scenario. First, although
$m_{c}$ ($\sim 1.3\,\mathrm{GeV}$) is larger than $\Lambda _{\mathrm{QCD}}$ (%
$\sim 0.2\mathrm{-}0.4\ \mathrm{GeV}$, depending on the number of effective
flavors), it is actually of the same order of magnitude as the nucleon mass,
which must certainly be considered as being of a nonperturbative scale.
Secondly, the ansatz itself is ill-defined, since: (i) the initial condition
$c(x,\mu _{0})=0$ depends sensitively on the choice of $\mu _{0}$; and (ii)
PQCD only suggests that $\mu _{0}$ be of the same order of magnitude as $%
m_{c}$, but does not dictate any particular choice. Since the $\mu $%
-dependence of $c(x,\mu )$ is relatively steep in the threshold region, the
condition $c(x,\mu _{0})=0$ for a given choice of $\mu _{0}$ is physically
quite different from that at a different choice of $\mu _{0}$. Thirdly, many
nonperturbative models give nonzero predictions for $c(x,\mu _{0})$---again,
for unspecified $\mu _{0}\sim m_{c}$ \cite{BrodskyHoyer,Brodsky,pumplinLC}.
We therefore wish to study the nucleon structure in a formalism that allows
for nonperturbative charm.

To carry out a systematic study of the charm sector of nucleon
structure, one needs: (i) a general global analysis framework that includes
a coherent treatment of nonzero quark masses in PQCD; and (ii) comprehensive
experimental inputs that have the potential to constrain the charm degrees
of freedom.\footnote{%
Earlier efforts \cite{HarrisSmithVogt,GunionVogt} treated light partons and
the IC component as dynamically uncoupled, and were done outside the
framework of a global analysis.} Recent advances on both fronts make this
study now possible. In the following, we extend the recent CTEQ6.5 global
analysis \cite{cteq65m,cteq65s} to include a charm sector with its own
independent degrees of freedom at the initial factorization scale $%
\mu_{0}=m_{c}$. We shall address the questions posed in the introduction by
examining the results of global analyses performed under three
representative scenarios.

The first two scenarios invoke the light-cone Fock space picture \cite%
{LightCone} of nucleon structure. In this picture, IC appears mainly at
large momentum fraction $x$, because states containing heavy quarks are
suppressed according to their off-shell distance, which is proportional to $%
(p_{\perp }^{\;2}+m^{2})/x$. Hence components with large mass $m$ appear
preferentially at large $x$. It has recently been shown that indeed a wide
variety of light-cone models all predict similar shapes in $x$ \cite%
{pumplinLC}.

The specific light-cone models we take as examples are the original model of
Brodsky et al.~\cite{BHPS} (BHPS), and a model in which the intrinsic charm
arises from virtual low-mass meson + baryon components such as $\overline{D}%
^{0}\Lambda_{c}^{+}$ of the proton---a \textquotedblleft meson
cloud\textquotedblright\ picture \cite{Navarra,Melnitchoukccbar,SteffensMT}.
It has been shown that these two models reasonably represent other natural
choices in the light-cone picture \cite{pumplinLC}. They are fairly similar
to each other as well; but the meson cloud model also predicts a difference
between $c(x)$ and $\bar{c}(x)$, so it provides an estimate of the size of
possible charm/anti-charm differences.

The light-cone formalism is not developed to a point where the normalization
of $uudc\bar{c}$ or $uudb\bar{b}$ components can be calculated with any
confidence \cite{pumplinLC}---though estimates on the order of $1\%$ have
been proposed in several studies \cite{Navarra,BagModel,Song}. In our
phenomenological approach, it is appropriate to treat the magnitude of IC as
a quantity to be determined by comparison with data.

The $x$-dependence predicted by the BHPS model~\cite{BHPS} is
\begin{equation}
c(x)=\bar{c}(x)=A\,x^{2} 
\left[ 6\,x\,(1+x)\,\ln x\,+\,(1-x)(1+10x+x^{2}) \right] .  \label{bhps}
\end{equation}
We use this form at scale $\mu _{0}=m_{c}$ for the BHPS scenario. The
normalization constant $A$ that controls the magnitude of IC in this model is
treated as a variable parameter in our study. 
The magnitude of IC can be conveniently characterized by the momentum 
fraction carried by $c + \bar{c}$---see Eq.~(\ref{magnitude}).

The exact $x$-dependence predicted by the meson cloud model cannot be
given by a simple formula, since it involves convolutions of the charmed
meson and baryon distributions in the proton with the quark distributions
within the meson and baryon. However, it was shown in Ref.~\cite{pumplinLC}
that the charm distributions in this model can be very well approximated by
\begin{eqnarray}
c(x) &=&A\,\,x^{1.897}\,(1-x)^{6.095}  \label{m-c1} \\
\bar{c}(x) &=&\bar{A}\,\,x^{2.511}\,(1-x)^{4.929}  \label{m-c2}
\end{eqnarray}
where the normalization constants $A$ and $\bar{A}$ are determined by
the requirement of the quark number sum rule 
$\int_{0}^{1}[c(x)-\bar{c}(x)]\,dx=0$, which specifies $A/\bar{A}$; and 
the overall magnitude of IC that is to be varied in our study.

In contrast to these light-cone scenarios, we also examine a purely
phenomenological scenario in which the shape of the charm distribution is
\emph{sea-like}---i.e., similar to that of the light flavor sea quarks,
except for an overall mass-suppression. In this scenario, for
simplicity, we assume $c(x)=\bar{c}(x)\propto \bar{d}(x)+\bar{u}(x)$ at the 
starting scale $\mu _{0}=m_{c}$.

In each of the three scenarios, the initial nonperturbative $c(x)$ and 
$\bar{c}(x)$ specified above is used as input to the general-mass perturbative 
QCD evolution framework discussed in detail in \cite{cteq65m}.  We then 
determine the range of magnitudes for IC that is consistent with our standard 
global analysis fit to data. This is described in the next Section.

%TCIDATA{Version=4.10.0.2363}
%TCIDATA{LaTeXparent=0,0,../charm.tex}

%TCIDATA{ChildDefaults=chapter:3,page:1}

\section{Global QCD Analysis with Intrinsic Charm}

\label{sec:GlobalChisqr}

For these global fits, we use the same theoretical framework and
experimental input data sets that were used for the recent CTEQ6.5 analysis
\cite{cteq65m}, which made the traditional ansatz of no IC. Notable
improvements over previous CTEQ global analyses \cite{cteq6} are: (i) the
theory includes a comprehensive treatment of quark mass effects in DIS
according to the PQCD formalism of Collins \cite{Collins}; and (ii) the full
HERA I data on Neutral Current and Charged Current total inclusive cross 
sections, as well as heavy
quark production cross sections, are incorporated. Published correlated
systematic errors are used wherever they are available.  Fixed-target DIS, 
Drell-Yan, and hadron collider data that were used previously are also 
included \cite{cteq6,cteq65m}.

For each model of IC described in Sec.~\ref{sec:IntrinsicCharm}, we carry
out a series of global fits with varying magnitudes of the IC component. From
the results, we infer the ranges of the amount of IC allowed by current data
within each scenario.

It is natural to characterize the magnitude of IC by the momentum fraction 
$\langle x\rangle _{c+\bar{c}}$ carried by charm at our starting scale for
evolution $\mu =1.3\,\mathrm{GeV}$. This is just the first moment of the 
$c + \bar{c}$ momentum distribution:\footnote{%
In the context of light-cone models, the zeroth moment $\int_{0}^{1} c(x) \,
dx = \int_{0}^{1} \bar{c}(x) \, dx$ (= total number of charm pairs) is
often used to characterize the magnitude of IC. In more general cases, it 
is not a good measure of IC because it is overly sensitive to the small-$x$
behavior. In particular, this integral does not even converge 
if $c(x)$ has the same power-law behavior as light quarks in the limit 
$x \to 0$, as in our sea-like model.}
\begin{equation}
\langle x\rangle _{c+\bar{c}}=\int_{0}^{1} x\left[ c(x)+\bar{c}(x)\right]
\,dx \; .  \label{magnitude}
\end{equation}

The quality of each global fit is measured by a global $\chi_{\mathrm{global}%
}^{2}$, supplemented by considerations of the goodness-of-fit to the
individual experiments included in the fit. (The procedure has been fully
described in \cite{cteq6,cteq65m}.) The three curves in Fig.~\ref{fig:figA}
show $\chi _{\mathrm{global}}^{2}$ as a function of $\langle x\rangle
_{c+\bar{c}}$ for the three models under consideration.%
\figA

We observe first that in the lower range, $0<\langle x\rangle
_{c+\bar{c}}\sim 0.01$, $\chi _{\mathrm{global}}^{2}$ varies very
little, i.e., the quality of the fit is very insensitive to 
$\langle x\rangle _{c+\bar{c}}$ in this interval. This means that 
\emph{the global analysis of hard-scattering data provides no evidence 
either for or against IC up to }$\langle x\rangle _{c+\bar{c}}\sim 0.01$.

Beyond $\langle x\rangle _{c+\bar{c}}\sim 0.01$, all three curves in Fig.~%
\ref{fig:figA} rise steeply with $\langle x\rangle _{c+\bar{c}}$---global
fits do place useful upper bounds on IC\,! The upper dots along the three
curves represent marginal fits in each respective scenario, beyond which 
the quality of the fit becomes unacceptable according to the procedure
established in Refs.\thinspace \cite{cteq6,cteq65m}---one or more of the
individual experiments in the global fit is no longer fitted
within the 90\% confidence level.\footnote{%
There is no particular data set that can be identified as particularly
sensitive to IC---the limiting feature is generally one of the
high-precision large-statistics DIS experiments from HERA.} This implies,
\emph{the global QCD analysis rules out the possibility of an IC component
much larger than }$0.02$\emph{\ in momentum fraction}.

We note that the allowed range for $\langle x\rangle_{c+\bar{c}}$ is somewhat
wider for the sea-like IC model. This is understandable, because under this
scenario, the charm component is more easily interchangeable with the other
sea quark components, in its contribution to the inclusive cross sections that
are used in the analysis; whereas the hard $c$ and $\bar{c}$ components of
the light-cone models are not easily mimicked by other sea quarks.

The PDF sets that correspond to the three limiting cases (upper dots), along
with three lower ones on the same curves that represent typical, more
moderate, model candidates (lower dots), will be explored in detail next.

\paragraph{BHPS model results:}

Figure~\ref{fig:figB} shows the charm distributions $c(x) = \bar{c}(x)$ at
three factorization scales that arise from the BHPS model, along with results
from the CTEQ6.5 PDFs which have no IC.%
\figB %
The short-dash curves correspond to the marginally allowed amount of IC
($\langle x\rangle _{c+\bar{c}}=0.020$) indicated by the upper dot on the
BHPS curve in Fig.~\ref{fig:figA}. The long-dash curves correspond to IC that
is weaker by a factor of $\sim 3$, indicated by the lower dot ($\langle
x\rangle _{c+\bar{c}}=0.0057$) on the BHPS curve in Fig.~\ref{fig:figA}. This
point corresponds to the traditional estimate of $1\%$ IC probability
in the BHPS model, i.e., $\int_{0}^{1}c(x)\,dx=\int_{0}^{1}\bar{c}%
(x)\,dx=0.01$, at the starting scale $\mu _{0}=1.3\,\mathrm{GeV}$. \emph{%
This physically motivated light-cone model estimate thus lies well within
the phenomenological bounds set by our global analysis.}

We see that at low factorization scales, this model produces a peak in the
charm distribution at $x \approx 0.3$. This peak survives in the form of a
shoulder even at a scale as large as $\mu=100\,\mathrm{GeV}$. At that scale,
IC strongly increases $c(x)$ and $\bar{c}(x)$ above the gluon splitting
contribution at $x > 0.1$, while making a negligible contribution at 
$x < 0.1$.

\paragraph{Meson Cloud model results:}

Figure~\ref{fig:figC} shows the charm distributions that arise from the $%
D_{0}\,\Lambda _{c}^{+}$ meson cloud model, together with the results from
CTEQ6.5 which has no IC. In the meson cloud model, the charm ($c(x)$) and
anti-charm ($\bar{c}(x)$) distributions are different.\figC

The short-dash (short-dash-dot) curves correspond to the maximum amount of
IC $c(x)$ ($\bar{c}(x)$) that is allowed by the data ($\langle x\rangle_{c+%
\bar{c}} = 0.018$), while the long-dash (long-dash-dot) curves show a
smaller amount ($\langle x\rangle_{c+\bar{c}} = 0.0096$), and the shaded
region shows CTEQ6.5 which has no IC. Again we see that IC can substantially
increase the charm PDFs at $x > 0.1$, even at a large factorization scale.

The difference between $c$ and $\bar{c}$ due to IC is seen to be potentially
quite large. The sign of the difference is such that $\bar{c}(x)>c(x)$ for $%
x \to 1$, as explained in Ref.~\cite{pumplinLC}. Experimental evidence for $%
c(x) \neq \bar{c}(x)$ may be difficult to obtain; but it is worth
considering because it would provide an important constraint on the
nonperturbative physics---as well as supplying a direct proof of intrinsic
charm, since $c\bar{c}$ pairs produced by gluon splitting are symmetric to
NLO.\footnote{%
QCD evolution at NNLO can produce $c(x) \neq \bar{c}(x)$ from the valence $%
u(x) \neq \bar{u}(x)$ and $d(x) \neq \bar{d}(x)$, but the predicted
asymmetry is only a few percent \cite{ccbarNNLO}.}

\paragraph{Sea-like model results:}

Figure~\ref{fig:figD} shows the charm distributions that arise from a model
in which IC is assumed to have the same shape in $x$ as the light-quark sea $%
\bar{u}(x)+\bar{d}(x)$ at the starting scale $\mu _{0}=1.3\,\mathrm{GeV}$.
The short-dash curve again corresponds to the maximum amount of IC of this
type that is allowed by the data ($\langle x \rangle_{c+\bar{c}}=0.024$),
while the long-dash curve shows an intermediate amount ($\langle x\rangle
_{c+\bar{c}}=0.011$), and the shaded region shows CTEQ6.5 which has no IC.%
\figD

We see that IC in this sea-like form can increase the charm PDFs over a
rather large region of moderate $x$. As a result, it may have important
phenomenological consequences for hard processes that are initiated by charm.

%TCIDATA{Version=4.10.0.2363}
%TCIDATA{LaTeXparent=0,0,../charm.tex}

%TCIDATA{ChildDefaults=chapter:4,page:1}

\section{Comparison with Light Partons}

In this section, we compare the charm content of the nucleon with the other
flavors at various hardness scales. Figure~\ref{fig:figCOMPcx} shows the charm
distribution with no IC, or IC with shape given by the BHPS model with the
two strengths discussed previously, compared to gluon, light quark, and light
antiquark distributions from the CTEQ6.5 best fit. By comparing the three
panels of this figure, one can recognize the standard characteristics of
DGLAP evolution:  at increasing scale, the PDFs grow larger at small $x$ and
smaller at large $x$; while the differences between $q$ and $\bar{q}$, and the
differences between flavors, all get smaller.

This figure shows that the light-cone form for IC has negligible effects
for $x < 0.05$, while it can make $c(x)$ and $\bar{c}(x)$ larger than any
of $\bar{u}(x)$, $\bar{d}(x)$, $s(x)$, and $\bar{s}(x)$ for $x > 0.2\,$.
Similar results hold for the meson-cloud model (not shown), since the
essential basis is the large-$x$ behavior that is characteristic of the
light-cone picture.%
\figCOMPcx %

\figCOMPca
Figure~\ref{fig:figCOMPca} similarly shows the charm content for
the scenarios of no IC, or IC with a ``sea-like'' shape for the two
normalizations discussed previously. In this case, unlike the light-cone
forms, charm remains smaller than the other sea quarks, including $s$ and
$\bar{s}$, at all values of $x$. However, this figure shows that IC can 
raise the $c$ and $\bar{c}$ distributions by up to a factor of $\sim 2$ 
above their traditional radiative-only estimates, so it can have an 
important effect on processes that are initiated by charm.

Figures~\ref{fig:figCOMPcx}--\ref{fig:figCOMPca} also show that in every
scenario, $c(x)$ and $\bar{c}(x)$ remain small compared to
$u$, $d$, and $g$.

%TCIDATA{Version=4.10.0.2363}
%TCIDATA{LaTeXparent=0,0,../charm.tex}

%TCIDATA{ChildDefaults=chapter:5,page:1}

\section{The Charm Distribution at Various Energy Scales}

\label{sec:ChmPdfs}

The effect of evolution of the charm distributions in each
scenario can be seen by comparing the three panels within each of
Figs.~\ref{fig:figB}--\ref{fig:figD}.  But to show the evolution
in more detail, we plot the $x$-dependence at scales
$\mu =1.3$, $2.0$, $3.16$, $5$, $20$, $100 \,\mathrm{GeV}$ together
on the same figure.  In place of the logarithmic scale in $x$, we
use a scale that is linear in $x^{1/3}$ in order to display the
important large-$x$ region more clearly.

\figE

The left panel of Fig.~\ref{fig:figE} has no IC; the center panel is the BHPS
model (maximal level); and the right panel is the sea-like IC model (maximal
level). In all cases, as the scale increases, the charm distribution becomes
increasingly soft as the result of QCD evolution. As one would expect, the IC
component dominates at low energy scales. The radiatively generated component
(coming mainly from gluon splitting) increases rapidly in importance with
increasing scale. A two-component structure appears in these plots at
intermediate scales, say $\mu =5 \, \mathrm{GeV}$, where IC is dominant at
large $x$ and radiatively generated is dominant at small $x$.  As noted
previously, even at a rather high energy scale of $\mu =100 \,\mathrm{GeV} $
the IC component is still very much noticeable in the light-cone wave
function models.  In all cases, the intrinsic component is quite large in
magnitude compared to the purely radiatively generated case in the region,
say, $x > 0.1 \,$. Thus, the existence of IC would have observable
consequences in physical processes at future hadron colliders that depend on
the charm PDF in this kinematic region.

%TCIDATA{Version=4.10.0.2363}
%TCIDATA{LaTeXparent=0,0,../charm.tex}

%TCIDATA{ChildDefaults=chapter:6,page:1}

\section{Summary and Implications}

\label{sec:Conclusion}

As a natural extension of the CTEQ6.5 global analysis
\cite{cteq65m,cteq65s}, we have determined the range of magnitudes for 
intrinsic charm that is consistent with an up-to-date global QCD analysis of
hard-scattering data, for various plausible assumptions on the shape of the 
$x$-distribution of IC at a low factorization scale.

For shapes suggested by light-cone models, we find that the global analysis
is consistent with anywhere from zero IC up to $\sim \! 3$ times the amount 
that has been estimated in more model-dependent studies.  In these models, 
there can be a large enhancement of $c(x)$ and $\bar{c}(x)$ at $x > 0.1$, 
relative to previous PDF analyses which assume no IC. The enhancement 
persists to scales as large as $\mu \sim 100 \, \mathrm{GeV}$, and can 
therefore have an important influence on charm-initiated processes at the 
Tevatron and LHC.  Large differences between $c(x)$ and $\bar{c}(x)$ for 
$x > 0.1$ are also natural in some models of this type.

For an assumed shape of IC similar to other sea quarks at low factorization
scale, there can also be a significant enhancement of $c(x)$ and $\bar{c}(x)$
relative to traditional no-IC analyses. In this case, the enhancement is spread
more broadly in $x$, roughly over the region $0.01 < x < 0.50$.

What experimental data could be used to pin down the intrinsic charm
contribution? An obvious candidate would be $c$ and $b$ production data from
HERA; but these are already included in the analysis, and they don't have
very much effect because the errors are rather large and because the data are
mostly at small $x$ where the dominant partonic subprocess is $\gamma g \to c
\bar{c}$ rather than $\gamma c \to c X$. It may be possible to probe the
subprocess $\gamma c \to c X$ more effectively by measuring specific angular
differential distributions---see \cite{Ananikyan}.

Future hadron collider measurements could place direct constraints
on the charm PDF. For instance, in principle, the partonic process
$g + c \to \gamma/Z+c$ is directly sensitive to the initial state
charm distribution. The experimental signature would be $Z$ plus a
tagged charm jet in the final state. Such measurements would be
experimentally very challenging, but potentially important
\cite{TeV4LhcQcd,cteq65impact}. The proposed future $ep$ colliders eRHIC
\cite{eRHIC} and LHeC \cite{LHeC} would also be very helpful for the
study of heavy quark parton distributions.

If IC is indeed present in the proton at a level on the order of what is
allowed by the current data, it will have observable consequences in physical
processes at future hadron colliders that depend on the charm PDF in the
large $x$ region.  Interesting examples that have been proposed in the
literature are: diffractive production of neutral Higgs bosons
\cite{Brodsky:2006wb}, and charged Higgs production at the LHC
\cite{HeYuan,TeV4LhcHiggs}. Application of our results to the latter process
will be presented in \cite{cteq65s}.

The PDF sets with intrinsic charm that were used in this paper will be made
available on the CTEQ webpage and via the LHAPDF standard, for use in
predicting/analyzing experiments.  They are designated by CTEQ6.5C.

\paragraph{Acknowledgment}

We thank Joey Huston, Pavel Nadolsky, Daniel Stump, and C.-P. Yuan for useful
discussions. This work was supported in part by the U.S. National Science
Foundation under award PHY-0354838, and by National Science Council of
Taiwan under grant 94(95)-2112-M-133-001.

\end{document}